# Diagnostic Uncertainty in Pneumonia Detection using CNN MobileNetV2 and CNN from Scratch

Kennard Norbert Sudiardjo
School of Computer Science
Bina Nusantara University
Jakarta, Indonesia 11480
kennard.sudiardjo@binus.ac.id

Islam Nur Alam
School of Computer Science
Bina Nusantara University
Jakarta, Indonesia 11480
islam.alam@binus.ac.id

Wilson Wijaya
School of Computer Science
Bina Nusantara University
Jakarta, Indonesia 11480
wilson.wijaya002@binus.ac.id

Lili Ayu Wulandhari
School of Computer Science
Bina Nusantara University
Jakarta, Indonesia 11480
lili.wulandhari@binus.ac.id

*Abstract— Pneumonia Diagnosis, though it is crucial for an effective treatment, it can be hampered by uncertainty. This uncertainty starts to arise due to some factors like atypical presentations, limitations of diagnostic tools such as chest X-rays, and the presence of co-existing respiratory conditions. This research proposes one of the supervised learning methods, CNN. Using MobileNetV2 as the pre-trained one with ResNet101V2 architecture and using Keras API as the built from scratch model, for identifying lung diseases especially pneumonia. The datasets used in this research were obtained from the website through Kaggle. The result shows that by implementing CNN MobileNetV2 and CNN from scratch the result is promising. While validating data, MobileNetV2 performs with stability and minimal overfitting, while the training accuracy increased to 84.87% later it slightly decreased to 78.95%, with increasing validation loss from 0.499 to 0.6345. Nonetheless, MobileNetV2 is more stable. Although it takes more time to train each epoch. Meanwhile, after the 10th epoch, the Scratch model displayed more instability and overfitting despite having higher validation accuracy, training accuracy decreased significantly to 78.12% and the validation loss increased from 0.5698 to 1.1809. With these results, ResNet101V2 offers stability, and the Scratch model offers high accuracy.*

## I. Introduction

Pneumonia, a type of infection that affects the lung tissue, is a significant global health concern. Especially in America, Asia, and Africa. There are evidence in the medical world about the challenges of the long term health in pneumonia survivors. Still, the path to diagnosing this condition isn't always uncomplicated. A layer of uncertainty can often linger, leading to an issue: misdiagnosis[1]. This introduction explores the complexities surrounding the diagnosis of pneumonia and the aspects that contribute to misleading the results. Despite the availability of diagnostic equipment like chest X-rays or blood tests, these procedures can be inaccurate. There are still no permanent regulations for diagnosing the symptoms of pneumonia. Rather, health caretakers took the conclusion based on the patient's symptoms, physical exam, or radiographic testing using chest X-rays. Still, experts say that this type of test, especially chest X-rays may lack sensitivity and can slow down the diagnosis process of the disease[2]. Similar evidence when the COVID-19 pandemic happened, where patients indicate symptoms that deviate from classic signs of pneumonia. Moreover, chest X-rays findings can be subtle, particularly in the early stages of pneumonia or they may imitate other respiratory illnesses like congestive heart failure or influenza[3]. Adding to the problem, some of the patients may already have an existing conditions that present with similar symptoms. This overlap can further mess up and making it difficult to see the difference between pneumonia and other diseases. Imagine a situation where a patient undergoes a set of tests, and based on the results, a diagnosis of pneumonia is confirmed. Antibiotics, the main treatment for ventilator-associated pneumonia, are prescribed. Still, the basic condition might not be pneumonia after all. This not only exposes the patient to unnecessary and potentially harmful drugs or medication but also delays the recovery of the true illness[4]. The rate of mortality in misdiagnosis of pneumonia, can cause interventions at the hospital where patients must prolong their duration of stay and run more unnecessary tests. These interventions could increase the number of viruses or bacteria that may lead to comorbidities where patients have more than one disease at the same time[5].

The research method that is going to be used is CNN MobileNetV2 and CNN from scratch. CNN MobileNetV2 was chosen because it's a highly effective deep learning method for image identification and classification. Its capability is important for machine vision, which integrate with recommender systems and linguistic communication process. CNN was designed to reduce process necessities, so that way it's more effective and easier to train data for image processing and linguistic communication processes[6]. Besides that, creating CNN from scratch using Keras library to be used as the second model because of the short training time and the low complexity[7].

## II. Related Works

Various AI learning such as machine learning have been used to develop techniques for detecting and classifying large sizes of datas. These techniques have been proven to assist researchers in achieving accurate outcomes, ultimately improving answers in the research. The research done by Alharbi et al., studying pneumonia x-ray images by implementing five pre-training Convolutional Neural Network(CNN) models and CAM metric for classification accuracy. Their research can detail the accuracy, sensitivity and specificity for the multiclassification of pneumonia[8]. One study by Liu et al., introduces a multi-branch fusion auxiliary learning (MBFAL) method for detecting pneumonia from chest x-ray images. Their research concludes that MBFAL is more effective as an auxiliary screening tool and outperforms other methods[9]. Another study by Sheu et al., developing a system that uses vital signs and chest x-ray images as an input by employing multimodal data analysis for pneumonia status prediction using deep learning classification (MDA-PSP). This system predicts whether pneumonia patients will be discharged from the hospital within 7 days or later, with hybrid methods ensuring higher prediction accuracy[10]. Another implementation by Szepesi et al., implemented a CNN model for accurate pneumonia detection from chest x-ray images. Instead of using pretrained networks and transfer learning, they created the CNN models from scratch, this concludes that the proposed model performs 97% better than its counter model based on the efficiency and accuracy[11]. CNN is an adaptive model in learning the hierarchy of features from existing data. The pre-trained model of CNN has been drilled with a large data set so that it can be used as a basis for more specific things. By removing the limitations of this model, it can produce a more effective and easier to train system for image processing.

## III. RESEARCH METHODOLOGY

### A. Datasets

The datasets used in this research were obtained from a website called "Kaggle" run and managed by Google, Inc, a collaborative website where people can find and publish datasets. The number of datasets about human diseases is widely available in Kaggle. This makes Kaggle a breakthrough for researchers as it provides secure access to many collections of datasets, which is crucial for training machine learning models. This research used datasets from a user by the name of "JTIPTJ", who provided the x-ray scan images of normal healthy lungs and the lung diseases, such as pneumonia, covid-19, and tuberculosis. The content of the datasets was converted into one table containing a list of the diseases and the value amount of the diseases in Table 1.

TABLE I. DATAEST TABLE.

| Table Names | Value |
|---|---|
| Normal | 1341 |
| Pneumonia | 3875 |
| Covid-19 Pneumonia | 460 |
| Tuberculosis | 650 |

Table 1 above shows the dataset containing the information about the quantity of every lung diseases on the list. The dataset contains 3 subfolders (train, test, and val) and each subfolder contains images from the category. This research used 6326 x-ray scan images in total as the primary source to differentiate and clustering every lung disease types available from the dataset. This research used python as the programming language to train the machine learning models for clustering the data and Kaggle as the source material of the datasets.

### B. CNN MobileNetV2

CNN is a supervised learning algorithm that is used to identify and classify images into a group based on the image type. CNN is designed to replicate neuron connections in the human brain, each neuron in the model scans the receptive area first to identify the easiest pattern before advancing to the more complex pattern. The technique in this method is by operating on the images based on patterns such as shapes and edges[12]

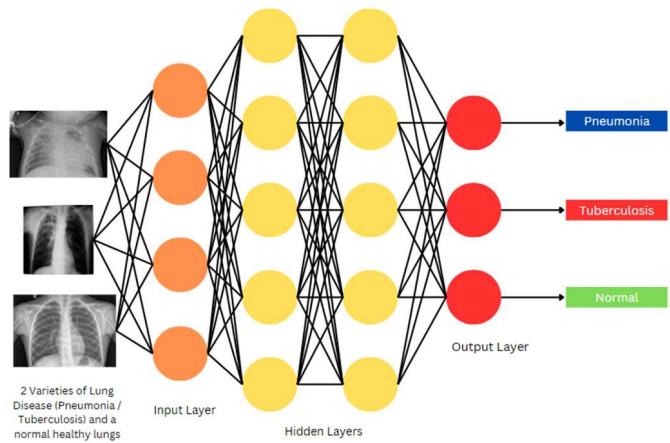

Fig 1. CNN implementation

MobileNetV2 is used as the pre-trained model; the main architecture for this model is ResNet101V2 because of the low memory usage, low expenses, and it only focused on the applications. An example from the image above is the process of comparing unhealthy and healthy lungs. CNN MobileNetV2 analyzes the data by identifying the simplest patterns first such as lines and shapes, before analyzing the more complex patterns such as lungs and other objects. Multiple hidden layers are present in the process, images are fed into the model to find the important parts. Lastly, the model used the learned information to receive what the image contains[13].

### C. CNN from scratch

CNN from scratch is similar to pre-trained CNN but creating it needs design and build for the entire architecture of the model as well as the type of layers and the numbers. The entire model is trained only with the specified dataset that is going to be used[7]. See Fig 1. Keras is used as the API for creating CNN from scratch because of the compact API, allowing to define complex CNN architectures with a few lines of code. Using Keras also allows the model to customize the activation functions, convolutional and pooling layers; this simplicity makes it more short and brief to experiment with different architectures[14]. Keras also helps the model to reuse layers and construct graphs, it also uses multi-GPU so that the model can run larger GPU clusters[15].

## IV. RESULT AND DISCUSSION

### A. Classifying using CNN MobileNetV2

## Training result

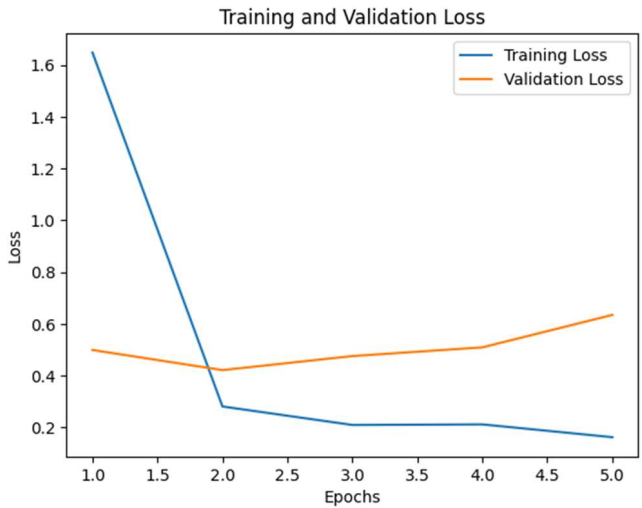

Fig 2. MobileNetV2 Training and Validation Loss Graph

At the start of the training, the model demonstrates strong performance with high training and validation accuracy, achieving 84.87% training accuracy and 81.58% validation accuracy. During the second epoch, there is a significant improvement in the training accuracy, reaching 91.59%, while the validation accuracy remains constant at 81.58%. This indicates that the model is starting to overfit, as it learns the training data too well without improving performance on the validation data. In the third epoch, the validation accuracy rises to 86.84%, showing that the model is starting to generalize better to previously unseen data. However, in the fourth epoch, validation accuracy drops to 78.95% and validation loss increases, indicating a decline inp performance on the validation set. By the fifth epoch, the training accuracy continues to improve, reaching 94.26%, but the validation accuracy remains stagnant at 81.58%, with increasing validation. This trend suggests that the model is continuing to overfit, improving on the training data but failing to generalize well to the validation data.

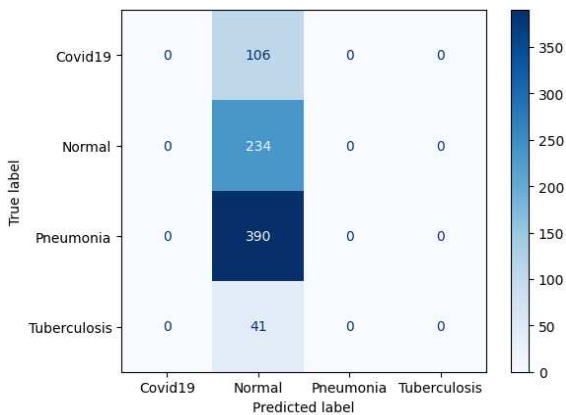

Fig 3. MobileNetV2 Confusion Matrix

For the result of the MobileNetV2 confusion matrix, it predicted 106 CXR images of covid-19, 234 CXR images of healthy lungs, 390 CXR images of pneumonia, and 41 CXR images of tuberculosis. Overall, the model shows strong initial performance with high accuracy at the first epoch and improvement in validation accuracy by the third epoch.

As evidenced for both training and validation, the model displays strong starting performance with high accuracy at the first epoch. At the third epoch there is an improvement in validation accuracy, but there was no further improvement. This implies that in order to avoid overfitting and enhance generalization, the model might need extra regulation methods like dropout, data augmentation, or extending the number of epochs with an early stopping strategy.

### B. Classifying using CNN from scratch

## Training result

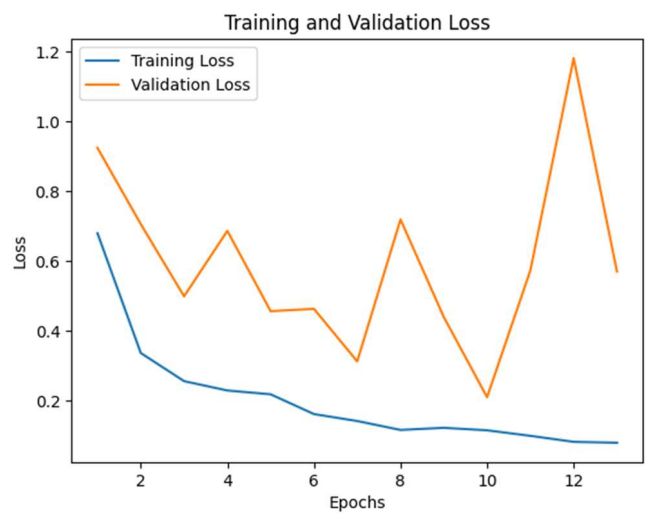

Fig 4. Scratch Training and Validation Loss Graph

In the first epoch, the model achieves a training accuracy of 73.24%, but the validation performance is still low, with an accuracy of 53.12%. By the second epoch, the model starts to learn and generalize better, with the training accuracy improving to 87.08% and the validation accuracy to 71.88%. In the third epoch, both training and validation accuracies increase further to 90.90% and 84.38%, respectively. However, in the fourth epoch, there is a slight decrease in validation accuracy and an increase in validation loss, indicating the start of overfitting. Despite this, by the fifth epoch, the model overcomes some of the overfitting, with the validation accuracy improving to 78.12%. In the sixth epoch, validation accuracy increases again to 84.38% and training accuracy reaches 94.38%, indicating that the model continues to overcome overfitting. The seventh epoch shows excellent validation performance with high accuracy and low loss. However, in the eighth epoch, validation loss increases sharply even though the accuracy remains constant at 84.38%, indicating potential overfitting. In the ninth epoch, validation loss decreases again, suggesting better generalization. By the tenth epoch, the validation accuracy peaks at 93.75%, while training accuracy reaches 96.46%, indicating that the model performs very well on the validation data. From the eleventh to thirteenth epochs, training accuracy remains high, ranging from 96.36% to 97.24%, but validation loss increases significantly and accuracy decreases to 78.12%, indicating that the model experiences serious overfitting during these final epochs.

For the result of the CNN scratch confusion matrix, it didn't show the result of the matrix. But, if comparing it to the MobileNetV2 the output would be likely the same due to the percentage similiarity of the model. See Fig 3.

The results indicate that the model overfits particularly after epoch 10, while training accuracy kept rising, validation accuracy fluctuated before declining altogether. The model performs remarkably well at first, improving in accuracy over training and validation data, until it reaches epoch 10. Applying additional regulation techniques like raising the dropout rate, augmentation of data, and L2 regulation is required to overcome overfitting. Early stopping should be used to stop training when validation accuracy is no longer increasing, this will stop additional overfitting. To achieve more optimal performance *Hyperparneedsers* such as learning rate, batch size, and model architecture (number of layers and neurons) need to be adjusted.

*C. Discussion*

MobileNetV2 demonstrates a strong inital performance with a high accuracy rate. This indicates that the model effectively captures the underlying patterns in the data from the start. However, the stagnant validation accuracy and increasing validation loss in later epochs suggest that the model is overfitting. Overfitting occurs when the model performs too well on training data but fails to generalize unseen data. This is a common issue in deep learning models, especially when the model has a high capacity to learn complex patterns in the training data. To prevent overfitting, techniques such as dropout, data augmentation, and early stopping can be implemented.

On the other hand, the CNN model trained form scratch shows a different learning curve compared to MobileNetV2. Initially the model struggles with lower validation accuracy, but it improves significantly by the third epoch, suggesting a steeper learning curve. However, the model starts to overfitting after the fourth epoch, it is indicated by the increasing validation loss. The validation accuracy peaks at the tenth epoch, but some epochs reveal a decline in performance due to overfitting. This suggests that while the CNN from scratch has a strong learning capacity, it also has a high tendency to memorize the training data rather than generalize from it. To improve the performance, regularization techniques such as increased dropout rates, data augmentation, and L2 regularization should be considered. Lastly, an early stopping mechanism could be beneficial to prevent model from training beyond the point of optimal generalization. Fine-tuning hyperparameters, such as learning rate, batch size, and the architecture of the model, could enhance performance.

When comparing both models, MobileNetV2 performs steadily and with minimal overfitting. Although it takes more time to train each epoch, this model is more stable. Meanwhile, after the tenth epoch, the Scratch model displays more severe instability and overfitting despite achieving a higher validation accuracy. In addition, MobileNetV2 takes approximately ~8000 seconds time per epoch compared with the Scratch model ~1500 seconds.

The choice between these models should be guided by the specific requirements of the tasks at hand. If the priority is to achieve the highest possible accuracy and there is a tolerance for overfitting, the CNN from scratch, with appropriate regularization, may be suitable. Conversely, if a more balanced performance with tability and less overfitting is desired, MobileNetV2 would bet he better option. In both cases, fine-tuning hyperparameters and employing advanced regularization techniques are essential steps to optimize model performance and generalization capability.

TABLE II. ACCURACY AND LOSS

| Model | Accuracy | | Loss | |
|---|---|---|---|---|
| | Training | Validation | Training | Validation |
| **MobileNetV2** | 84.87% - 94.26% | 81.58% - 86.84% | 0.1622 - 1.6475 | 0.4990 - 0.6345 |
| **Scratch** | 73.24% - 97.24% | 78.12% - 93.75% | 0.0793 - 0.6793 | 0.9236 - 1.1808 |

We can see from the table that both training accuracy and validation accuracy increased but in the Scratch model the validation accuracy decreased to 78.12% after the tenth epoch. Both model training losses decreased at the end of training meaning both models can minimize errors at training data. The validation loss at the MobileNetV2 model shows that this model is better at generalization than the model from scratch.

There are 2 models that we train to detect pneumonia using CNN deep learning method, MobileNetV2 model falls behind with the highest accuracy of 94.26% and Scratch model surpasses it with the accuracy of 97.24%. Compared to the other model trained by Mahir Kaya, 16 different models were individually trained to detect pneumonia using the transfer learning method. Among these, the top performers were DenseNet121, RegNetY008, and VGG19. RegNetY008 achieved the highest accuracy, scoring 95.19%[16].

V. CONCLUSION

Implementing CNN using MobileNetV2 and creating from scratch showed a promising result of classifying pneumonia based on the training accuracy. Nonetheless, it reveals that both models are struggling with overfitting. With the case of MobileNetV2, while the training accuracy increased steadily into 84.87%, validation accuracy remained stable but later decreased slightly in the other epochs from

81.58% into 78.95%. With the increasing validation loss from 0.499 to 0.6345, this shows that the model memorized all of the data but failed to memorize the unseen or invisible data. The CNN that is built from scratch at first shows good performance stepping up on both training and validation accuracy. Although, after the 10th epoch, it became overfitted and the validation loss was increased from 0.5698 to 1.1808 and accuracy decreased significantly to 78.12%. This is a warning as the model becomes more overly specific to the training examples and loses its ability to identify new images. For the most part, the CNN that was built from scratch achieved a higher validation accuracy compared to the CNN MobileNetV2, but its vulnerability to overfitting requires more attention. MobileNetV2 on the other hand, as the pre-trained model, has more potential by applying proper hyperparameter tuning and regularization techniques to reach the same or better results since it is computationally efficient from the added benefit as a pre-trained model.

## REFERENCES

[1]  Jones, B., & Waterer, G. (2020). Advances in community-acquired pneumonia. *Therapeutic advances in infectious disease*, *7*, 2049936120969607. https://journals.sagepub.com/doi/epdf/10.1177/2049936120969607?src=getftr

[2]  Metlay, J. P., Waterer, G. W., Long, A. C., Anzueto, A., Brozek, J., Crothers, K., ... & Whitney, C. G. (2019). Diagnosis and treatment of adults with community-acquired pneumonia. An official clinical practice guideline of the American Thoracic Society and Infectious Diseases Society of America. *American journal of respiratory and critical care medicine*, *200*(7), e45-e67. http://rimed.org/rimedicaljournal/2014/08/2014-08-20-cont-maughan.pdf

[3]  Cellina, M., Orsi, M., Toluian, T., Pittino, C. V., & Oliva, G. (2020). False negative chest X-Rays in patients affected by COVID-19 pneumonia and corresponding chest CT findings. *Radiography*, *26*(3), e189-e194. https://www.sciencedirect.com/science/article/pii/S1078817420300699?pes=vor

[4]  Fernando, A., Tran, A., Cheng, W., Klompas, M., Kyeremanteng, K., Mehta, S., ... & Rochwerg, B. (2020). Diagnosis of ventilator-associated pneumonia in critically ill adult patients—a systematic review and meta-analysis. *Intensive care medicine*, *46*, 1170-1179. https://link.springer.com/article/10.1007/s00134-020-06036-z

[5]  Hespanhol, V., & Bárbara, C. (2020). Pneumonia mortality, comorbities matter?. *Pulmonology*, *26*(3), 123-129. https://www.sciencedirect.com/science/article/pii/S2531043719302053

[6]  Chattopadhyay, A., & Maitra, M. (2022). MRI-based brain tumor image detection using CNN based deep learning method. *Neuroscience informatics*, *2*(4), 100060. https://www.sciencedirect.com/science/article/pii/S277252862200022X

[7]  El-Shafai, W., Mahmoud, A. A., Ali, A. M., El-Rabaie, E. S. M., Taha, T. E., El-Fishawy, A. S., ... & El-Samie, F. E. A. (2024). Efficient classification of different medical image multimodalities based on simple CNN architecture and augmentation algorithms. *Journal of Optics*, *53*(2), 775-787. https://www.researchgate.net/profile/Walid-El-Shafai/publication/366976747

[8]  Alharbi, A. H., & Hosni Mahmoud, H. A. (2022, May). Pneumonia transfer learning deep learning model from segmented X-rays. In *Healthcare* (Vol. 10, No. 6, p. 987). MDPI. https://www.mdpi.com/2227-9032/10/6/987

[9]  Liu, J., Qi, J., Chen, W., & Nian, Y. (2022). Multi-branch fusion auxiliary learning for the detection of pneumonia from chest X-ray images. *Computers in Biology and Medicine*, *147*, 105732. https://www.sciencedirect.com/science/article/pii/S0010482522005091

[10]  Sheu, R. K., Chen, L. C., Wu, C. L., Pardeshi, M. S., Pai, K. C., Huang, C. C., ... & Chen, W. C. (2022). Multi-modal data analysis for pneumonia status prediction using deep learning (MDA-PSP). *Diagnostics*, *12*(7), 1706. https://www.mdpi.com/2075-4418/12/7/1706

[11]  Szepesi, P., & Szilágyi, L. (2022). Detection of pneumonia using convolutional neural networks and deep learning. *Biocybernetics and biomedical engineering*, *42*(3), 1012-1022. https://www.sciencedirect.com/science/article/pii/S0208521622000742

[12]  Bhatt, D., Patel, C., Talsania, H., Patel, J., Vaghela, R., Pandya, S., ... & Ghayvat, H. (2021). CNN variants for computer vision: History, architecture, application, challenges and future scope. *Electronics*, *10*(20), 2470. https://www.mdpi.com/2079-9292/10/20/2470

[13]  Gulzar, Y. (2023). Fruit image classification model based on MobileNetV2 with deep transfer learning technique. *Sustainability*, *15*(3), 1906. https://www.mdpi.com/2071-1050/15/3/1906

[14]  Taparhudee, W., Jongjaraunsuk, R., Nimitkul, S., Suwannasing, P., & Mathurossuwan, W. (2024). Optimizing Convolutional Neural Networks, XGBoost, and Hybrid CNN-XGBoost for Precise Red Tilapia (Oreochromis niloticus Linn.) Weight Estimation in River Cage Culture with Aerial Imagery. *AgriEngineering*, *6*(2), 1235-1251. https://www.mdpi.com/2624-7402/6/2/70

[15]  Chicho, B. T., & Sallow, A. B. (2021). A comprehensive survey of deep learning models based on Keras framework. *Journal of Soft Computing and Data Mining*, *2*(2), 49-62. https://publisher.uthm.edu.my/ojs/index.php/jscdm/article/view/8732/4559

[16]  Kaya, M. (2024). Feature fusion-based ensemble CNN learning optimization for automated detection of pediatric pneumonia. *Biomedical Signal Processing and Control*, *87*, 105472. https://www.sciencedirect.com/science/article/abs/pii/S1746809423009059